\documentclass{iaus}
\usepackage{graphicx}

\title[PNe and the chemical evolution of the galactic bulge] %% give here short title %%
{Planetary nebulae and the chemical evolution of the galactic bulge: new abundances of older objects}

\author[O. Cavichia, R.D.D. Costa, M. Moll\'a, W.J. Maciel]   %% give here short author list %%
{Oscar Cavichia$^{1}$,
 Roberto D. D. Costa$^1$,
 Mercedes Moll\'a$^2$,
 \and Walter J. Maciel$^1$}

\affiliation{$^1$IAG, University of S\~ao Paulo, 05508-900, S\~ao Paulo-SP, Brazil.\\
$^{2}$CIEMAT, Universidad Complutense de Madrid, Spain.\\
email: cavichia@astro.iag.usp.br}

\pubyear{2011}
\volume{283}  %% insert here IAU Symposium No.
\jname{Planetary Nebulae: an Eye to the Future}
\editors{A. Manchado, L. Stanghellini \& D. Schoenberner, eds.}
\begin{document}

\maketitle

\begin{abstract}
In view of their nature, planetary nebulae have very short lifetimes, and the chemical abundances derived so far have a natural bias favoring younger objects. In this work, we report physical parameters and abundances for a sample of old PNe located in the galactic bulge, based on low dispersion spectroscopy secured at the SOAR telescope using the Goodman Spectrograph. The new data allow us to extend our database including older, weaker objects that are at the faint end of the planetary nebula luminosity function (PNLF). The results show that the abundances of our sample are lower than those from our previous work. Additionally, the average abundances of the galactic bulge do not follow the observed trend of the radial abundance gradient in the disk. These results are in agreement with a chemical evolution model for the Galaxy recently developed by our group.
\end{abstract}

\firstsection % if your document starts with a section,
\section{Introduction and Method}

Planetary nebulae (PNe) constitute an important tool to study the chemical evolution of the Galaxy, comprising the disk, bulge and halo populations (Cavichia et al. \cite{cavichia10,cavichia11}). Due to their nature, PNe have a very short lifetime, dissipating into the interstellar medium in a timescale of $10^4$ years. Therefore, the presently available observational results are strongly biased since they were focused on brighter and younger objects. 

In this work, we derived physical parameters and chemical abundances for 8 old and faint bulge PNe, making use of the size and optical quality of the 4.1m SOAR telescope located at Chile. These PNe were observed with the Goodman Spectrograph using a 300 l/mm grating, which covers all the optical domain. All observed PNe have high extinction with $E(B-V) > 2$ and low surface brightness. They are located in the direction of the galactic bulge, close to the galactic center. 
The observed sample was selected from Jacoby and Van de Steene (\cite{jacoby04}) and the MASH catalog (Parker et al. \cite{parker06}). Electron densities are obtained from $[\mbox{SII}]$ lines ratio: $671.6/673.0\mbox{ nm}$; and electron temperatures are calculated from the [NII] lines ratio: $(654.8+658.4)/575.5 \mbox{ nm}$. Due to the high extinction, in some cases the [NII] line 575.5 nm is not available in the spectrum. In this case, we estimated an upper limit for the 575.5 nm line flux. This is done using the background of the spectrum at this wavelength. In order to calculate the chemical abundances we used the empirical method described in Cavichia et al. (\cite{cavichia10}). 

Simulations predicted that radial gas flows are expected in barred galaxies (Athanassoula \cite{athanassoula92}, Friedli et al. \cite{friedli94}). In order to simulate the effects of the bar on the radial abundance gradient of the Galaxy, we updated the chemical evolution model of Moll\'a \& D\'iaz (\cite{molla05}), hereafter MD, including radial gas outflows with velocities of $1 \mbox{ km s}^{-1}$.
 
\section{Results and discussion}

\begin{figure}[!ht]
\begin{minipage}[t]{0.48\linewidth}
\centering
 \includegraphics[width=1.15\linewidth]{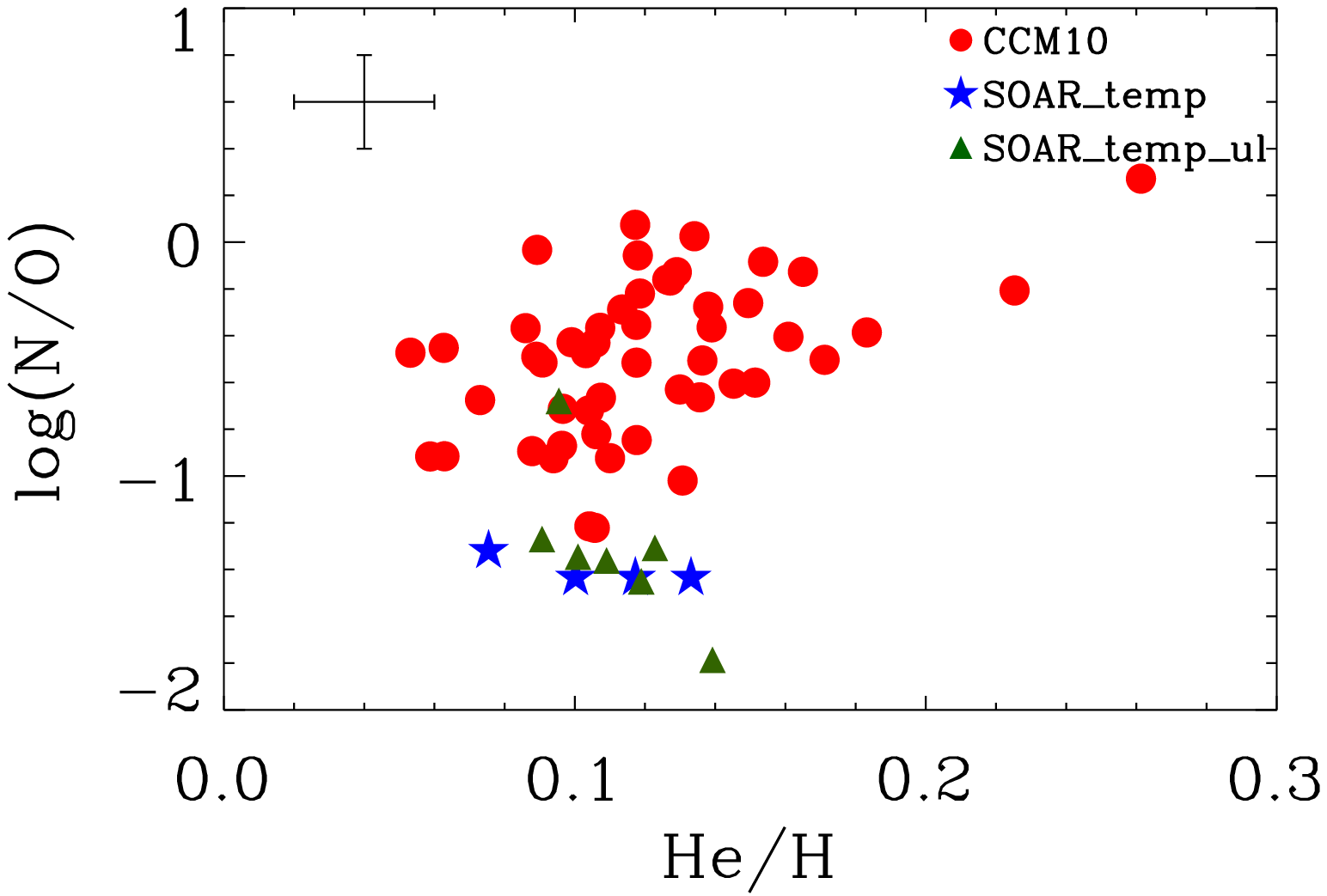}\\
 \caption{$\log (N/O)$ as a function of helium abundance for CCM10 (circles), the SOAR data using the traditional temperature method (stars), and the upper limit temperature method (triangles). The error bars at the top left are for CCM10 data.\label{logno}}
\end{minipage}
\hspace{0.3cm}
\begin{minipage}[t]{0.48\linewidth}
\centering
 \includegraphics[width=1.15\linewidth]{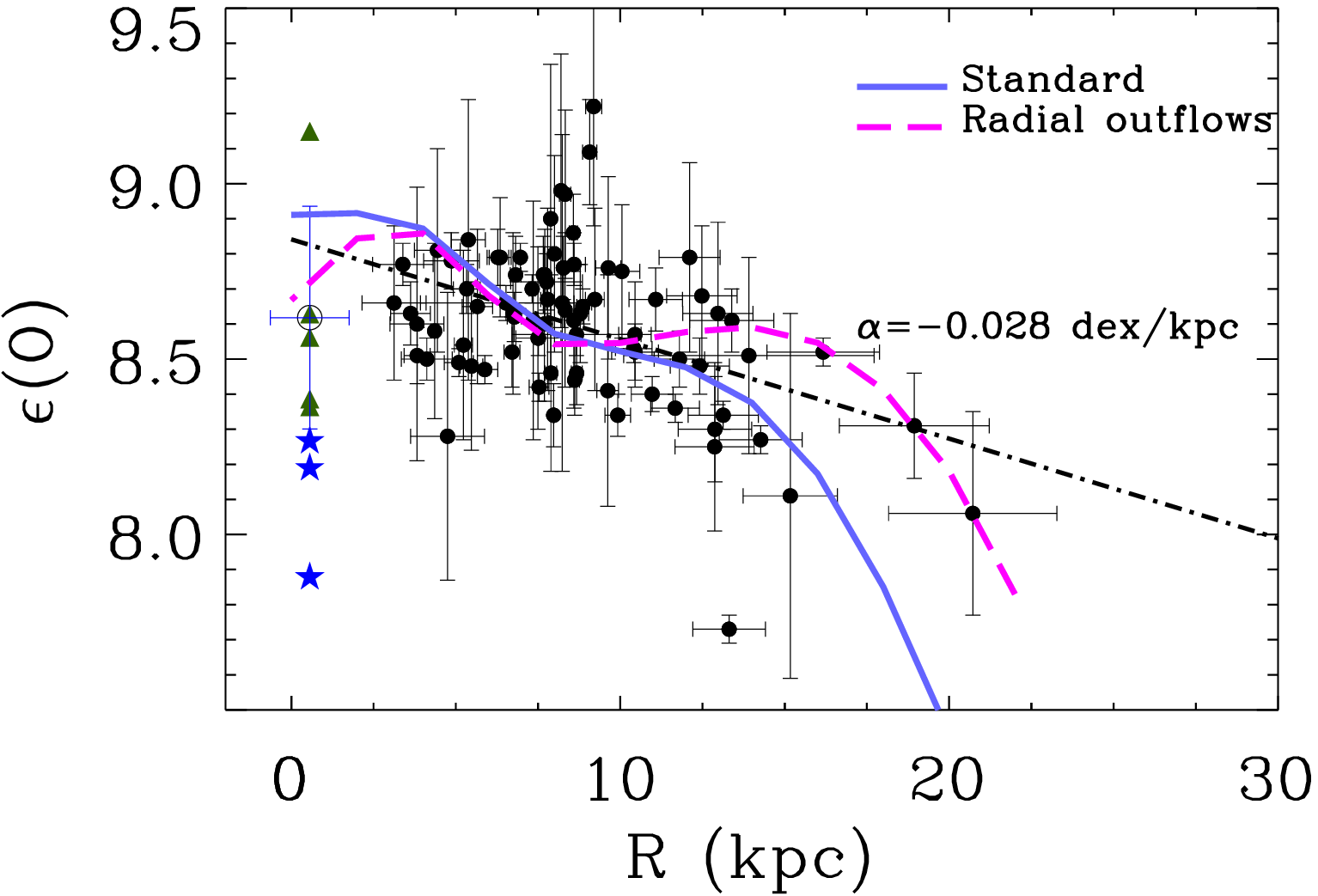}\\
 \vspace{-0.3cm}
 \caption{Oxygen radial gradient for SH data (circles). The unfilled circle is the mean abundance of our sample. The dash dotted line is a linear fit to the disk sample whose slope is indicated. The continuous line is the standard model from MD and the dashed line represents the model with radial gas outflows.\label{grad}}
\end{minipage}
\end{figure}

Figure \ref{logno} shows the $\log(\mbox{N}/\mbox{O})$ as a function of He/H abundances. Triangles and stars represent the data from this work, and filled circles the data from Cavichia et al. (\cite{cavichia10}), hereafter CCM10. Since PNe with high abundances of He and N are originated from massive stars, we expected that the PNe of our sample are originated from lower mass and, hence, older stars if compared with our previous work. 

The results from the chemical evolution model are shown in figure \ref{grad}. The model with radial gas outflows predict lower abundances in the inner region of the Galaxy and higher abundances in the outer region, when compared with the standard model from MD. These results are in agreement with the observational data for the disk from Stanghellini \& Haywood (\cite{stanghellini10}), hereafter SH, and with the observational data from the present work.

\vspace{-.9cm}

These results may have a significant impact on galactic evolution theories, providing a much more accurate view of the abundance distribution of bulge PNe, and therefore producing more reliable constraints for the modeling of intermediate mass stars evolution as well as the chemical evolution of the galactic disk and bulge.

{\small Acknowledgements. This work was partially supported by FAPESP (proc. 07/07704-2) and CAPES (proc. 3026/10-8).}

\end{document}